\documentclass[lettersize,journal]{IEEEtran}
\usepackage{amsmath,amsfonts}
\usepackage{algorithmic}
\usepackage{algorithm}
\usepackage{array}
\usepackage{booktabs} % 导入三线表需要的宏包
\usepackage{textcomp}
\usepackage{stfloats}
\usepackage{url}
\usepackage{verbatim}
\usepackage{graphicx}
\usepackage{cite}

\usepackage{graphicx} % 图片宏包
\usepackage{caption} % 主标题宏包
\usepackage{subcaption} % 子标题宏包
\usepackage{hyperref}
\begin{document}

\title{In-sensor Computing ANN Capacitive Sensors }

\author{Guihua Zhao\#, Yating Peng\#, Jiaxin Zhu\#, Xin Tang and Zhiyi Yu,~\IEEEmembership{Senior Member,~IEEE}
        % <-this % stops a space
\thanks{This work was supported in part by the Key-Area Research and Development Program of Guangdong Province under 2023B0303030004 and 2021B0101410004; in part by the National Natural Science Foundation of China (NSFC) under Grant 62334014.}% <-this % stops a space
\thanks{Guihua Zhao, Yating Peng, Jiaxin Zhu, Xin Tang and Zhiyi Yu are with School of Microelectronics Science and Technology, Sun Yat-sen University, Zhuhai 519082, China (zhaogh7@mail.sysu.edu.cn, pengyt8@mail2.sysu.edu.cn, zhujx33@mail2.sysu.edu.cn, 
tangx67@mail2.sysu.edu.cn, yuzhiyi@mail.sysu.edu.cn).}% <-this % stops a space
\thanks{Zhiyi Yu is also with Guangdong Provincial Key Laboratory of Optoelectronic Information Processing Chips and Systems, Sun Yat-Sen University, Guangdong, China.}% <-this % stops a space
\thanks{\#Guihua Zhao, Yating Peng and Jiaxin Zhu have equal contribution.}
\thanks{Manuscript submitted 27, 5, 2024.}}

\markboth{Journal of \LaTeX\ Class Files,~Vol.~, No.~27, May~2024}%
{Shell \MakeLowercase{\textit{et al.}}: A Sample Article Using IEEEtran.cls for IEEE Journals}


\maketitle

\begin{abstract}
This letter proposes an in-sensor computing multiply-and-accumulate (MAC) circuit based on capacitance. The MAC circuits can constitute an artificial neural network (ANN) layer and be operated as ANN classifiers and autoencoders. The proposed circuit is a promising scheme for capacitive ANN image sensors, showing competitively high efficiency and lower power.
\end{abstract}


\section{Introduction}

\IEEEPARstart{C}{apacitive} 
image sensors are used in various fields, such as  electronic skin, fingerprint identification, and gesture recognition.
In the conventional architecture, the capacitive sensors are physically separated from computing units. External signals like distance and pressure, are mapped to capacitive values, which are firstly converted to digital signals by analog-to-digital conversion (ADC) and then stored in memory before being sent to processing units. Due to sensor/processor interface and Von Neumann architecture, the conversion and transmission incur high latency and power consumption, which is a barrier to time-critical applications. To address this problem, we proposed an in-sensor computing paradigm of capacitve image sensors for ANNs. \cite{ref1, ref2, ref3, ref4, ref5, ref6}. The main contribution of this work includes: (1) a general capacitive in-sensor MAC circuit suitbable for ANN computing are proposed; (2) The algorithms including FC classifiers/autoencoders and CNN classifiers are presented and used to verify the proposed circuits. (3)The latency of correct recognition can reach 0.35$\mu$s for the proposed FC classifier.

\section{Capacitive in-sensor computing circuits}
\begin{figure}[h]
	\centering
	\includegraphics[width=0.4\textwidth]{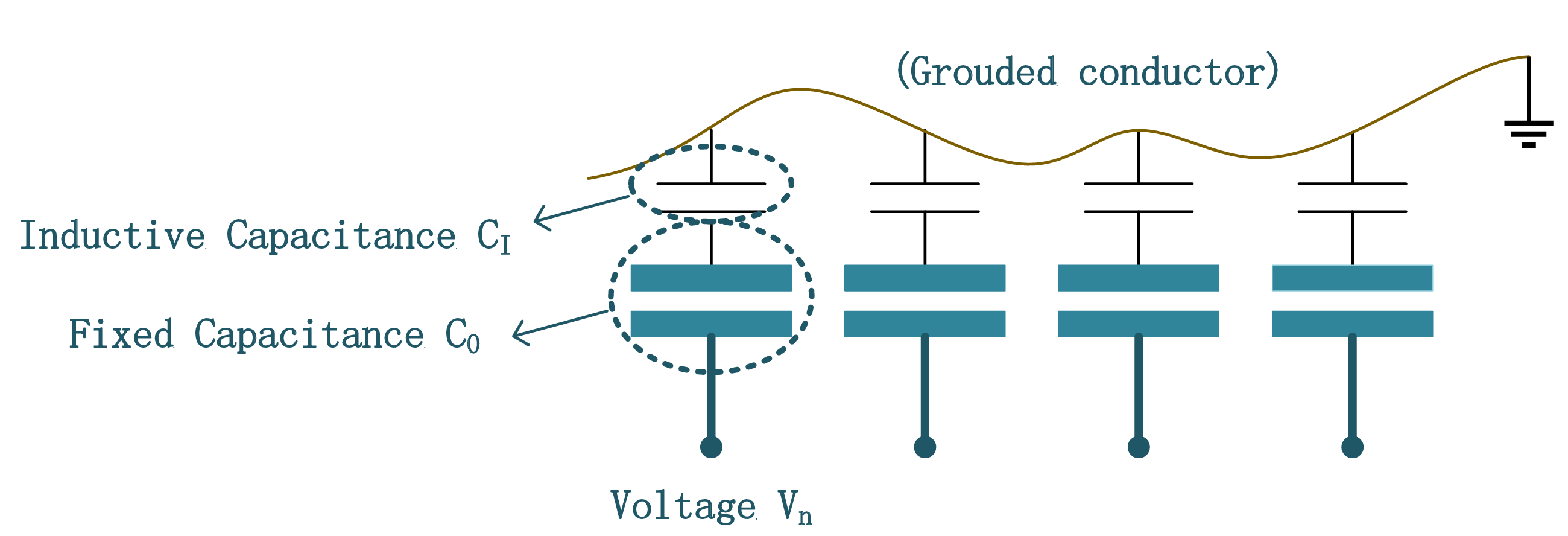}
	\caption{Schematic of inductive capacitance generation.} 
	\label{Fig1}
\end{figure}

A capacitive sensor of $ C_{0} $ is serially connected with an external capacity of $ C_{I} $ which depends on the distance between the upper electrode of $ C_{0} $  and the surface of grounded object,as shown in Fig.\ref{Fig1}. For the array, external capacitive pixels $ C_{I} $ form an image of object surface, and encoded to series of $ C_{0} $ and $ C_{I} $.
%\begin{figure}[h]
%	\centering
%	\includegraphics[width=0.5\textwidth]{MAC}
%	\caption{MAC} 
%	\label{MAC}
%\end{figure}
\begin{figure}[h]
	\centering
	\includegraphics[width=0.4\textwidth]{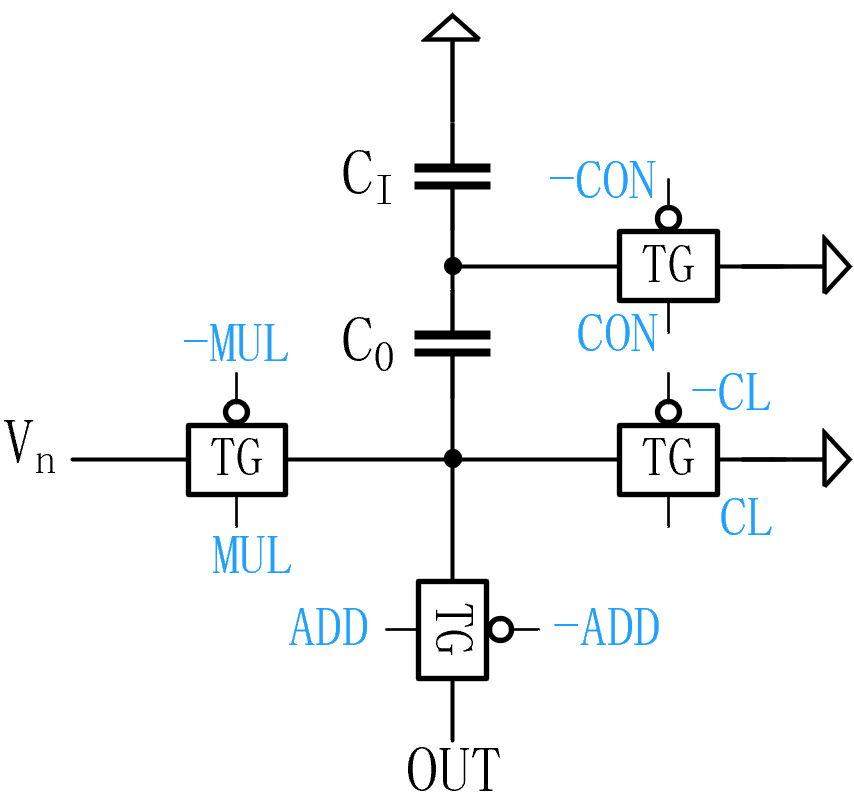}
	\caption{Schematic of a capacitive pixel.} 
	\label{Fig2}
\end{figure}

%\subsection{Capacitive in-sensor MAC circuit }
A capacitive in-sensor multiply-and-accumulate circuit is shown in Fig.\ref{Fig2}. Each unit of the sensor is controlled by four Transmission Gates. The MAC operation is segmented into four distinct phases: charge clearing, capacitor charging, charge transfer, and charge summation. Initially, the charge clearing step is executed by setting CL, CON, ADD high and MUL low, which clears any residual charge in the capacitor. In the second step, MUL is set high while CL, CON, and ADD remain low, charging the capacitor according to the input voltage $ V_{n} $ which serves as the weight. Consequently, the charge induced on the lower plate is $ Q_{n}=C_{n}×V_{n} $, where $C_{n} = ({C_{I,n} \times C_0})/({C_{I,n} + C_0})$. During the third step, CON is set high and MUL, CL, ADD are set low, stabilizing the charge while altering the capacitor to a fixed capacitance $C_{0} $, setting the voltage at the lower plate to $ U_{o,n}=Q_{n}/C_{0} $. Finally, with CON and ADD turned on and MUL, CL turned off, the charges from nine adjacent capacitors in parallel are summed and converted into the output voltage $U_{out}=\sum_{n=1}^{9}\left(C_{n}V_{n}\right)/9C_{0} $. Thus, this circuit serves as a fundamental unit for MAC operations.

\begin{figure*}[t!]
	\centering
	\begin{subfigure}[b]{0.42\textwidth} % 调整子图的宽度
		\centering
		\includegraphics[width=\textwidth]{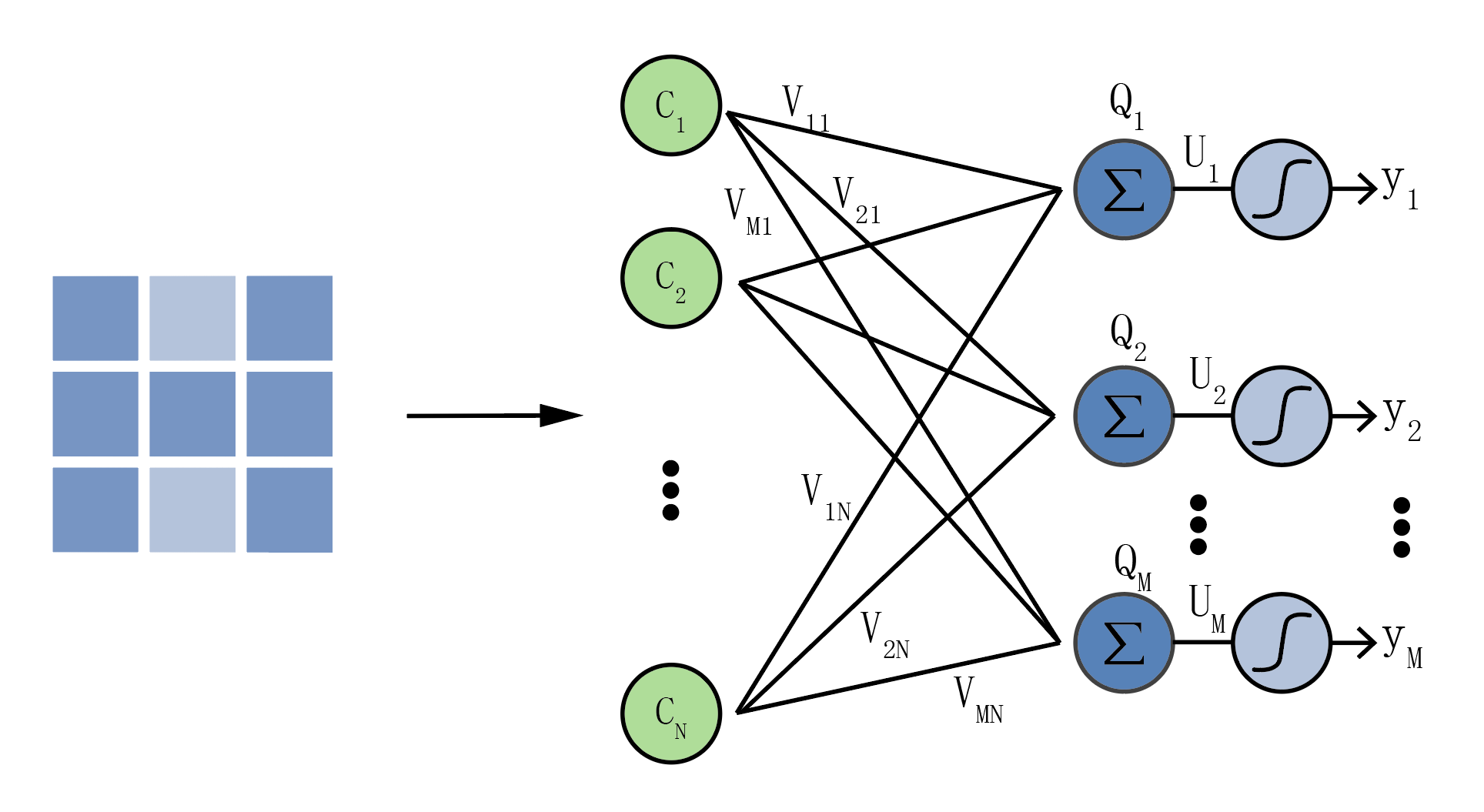} % 替换为您的图片文件名
		\caption{}
		\label{Fig3a}
	\end{subfigure}
	\hfill % 使用 \hfill 添加足够的空白以填充剩余的空间
	\begin{subfigure}[b]{0.57\textwidth} % 调整子图的宽度
		\centering
		\includegraphics[width=\textwidth]{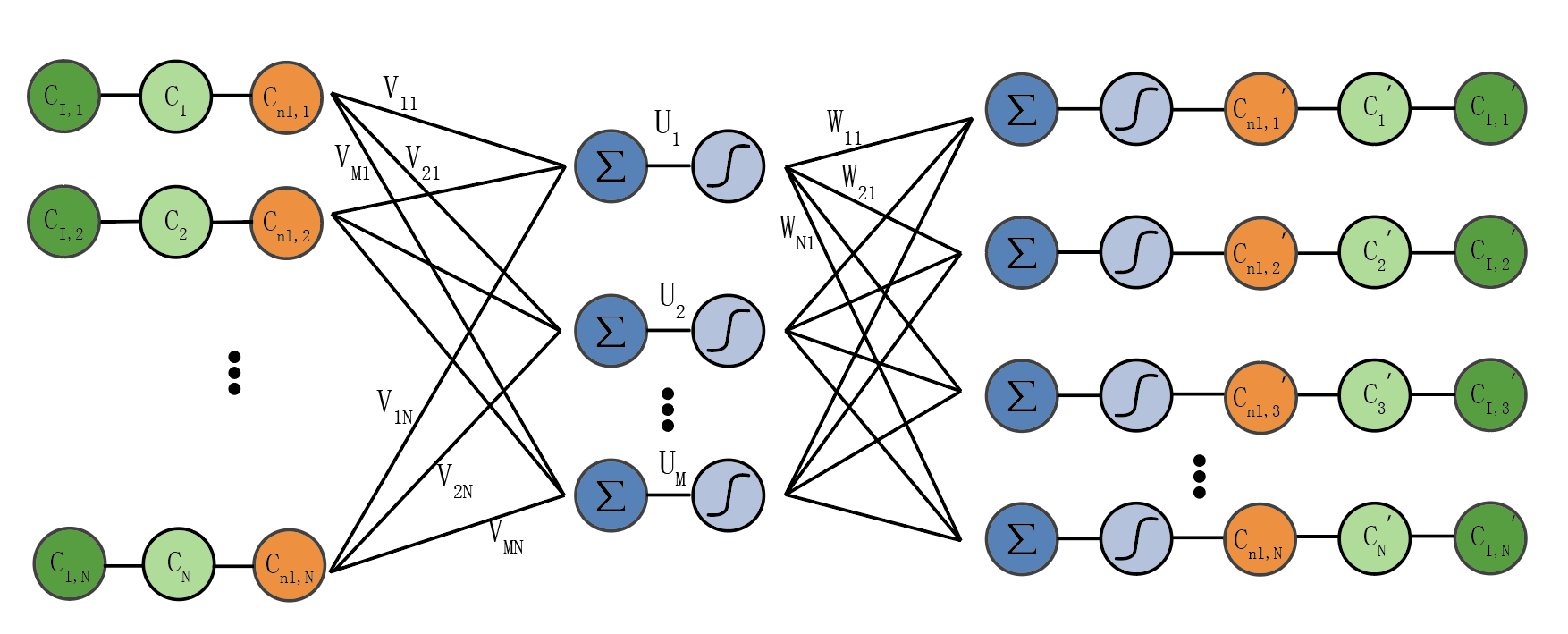} % 替换为您的图片文件名
		\caption{}
		\label{Fig3b}
	\end{subfigure}
	\caption{(a) Schematic of the FC classifiers and (b) Schematic of the FC autoencoder with computing of FC layers in the capacitive sensor array.}
	\label{Fig3}
\end{figure*}

\begin{figure*}[t!]
	\centering
	% 第一个 minipage 包含图1
	\begin{minipage}{0.48\textwidth}
		\centering
		\includegraphics[width=\textwidth]{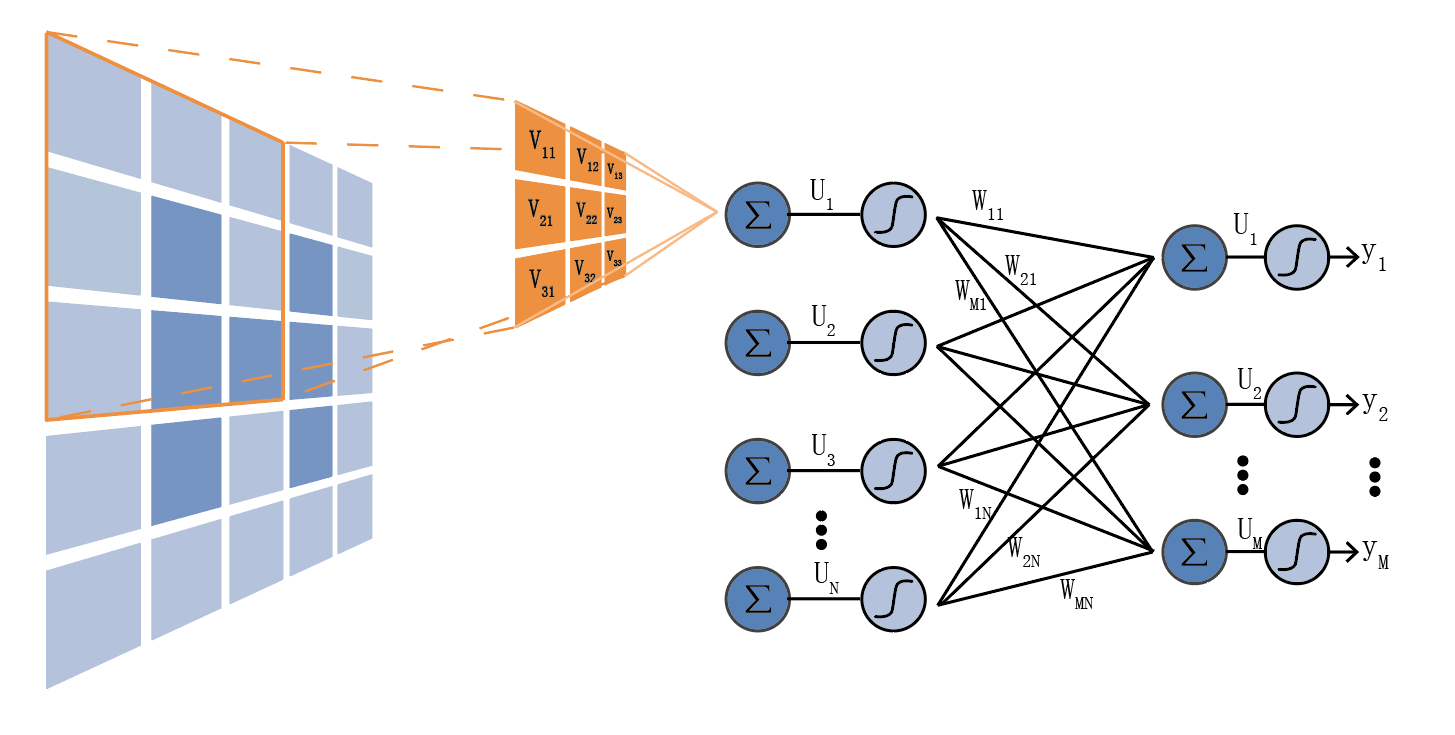}
		\caption{Schematic of the CNN classifiers with convolutional layer computing in the capacitive sensor array.}
		\label{Fig4}
	\end{minipage}
	% 第二个 minipage 包含图2和图3
	\hfill
	\begin{minipage}{0.5\textwidth}
		\centering
		\includegraphics[width=\textwidth]{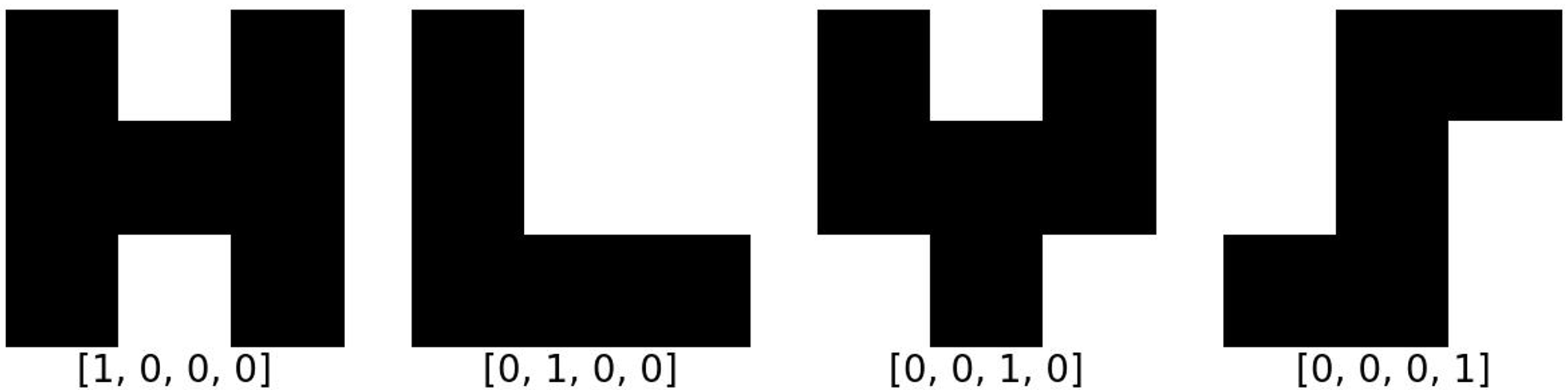}
	%	\subfloat[]{\includegraphics[width=1\textwidth]{Fig5}\label{Fig5}}
	%	\hfill
	%	\subfloat[]{\includegraphics[width=1\textwidth]{Fig14}\label{Fig14}}
		\caption{3 × 3 Pixel Images and Labels of the Letters H, L, Y, and Inverted Z for Neural Network Training.}
		\label{Fig5}
	\end{minipage}%
\end{figure*}

\section{Capacitive in-sensor computing systems for ANNs}
Capacitive sensor arrays can act as basic components for constructing artificial neural networks, used to perform the computational operations of fully connected and convolutional layers, as shown in Fig.\ref{Fig3} and \ref{Fig4}.

\begin{figure*}[h]
	\centering
	\begin{subfigure}{0.9\columnwidth}
		\includegraphics[width=\linewidth]{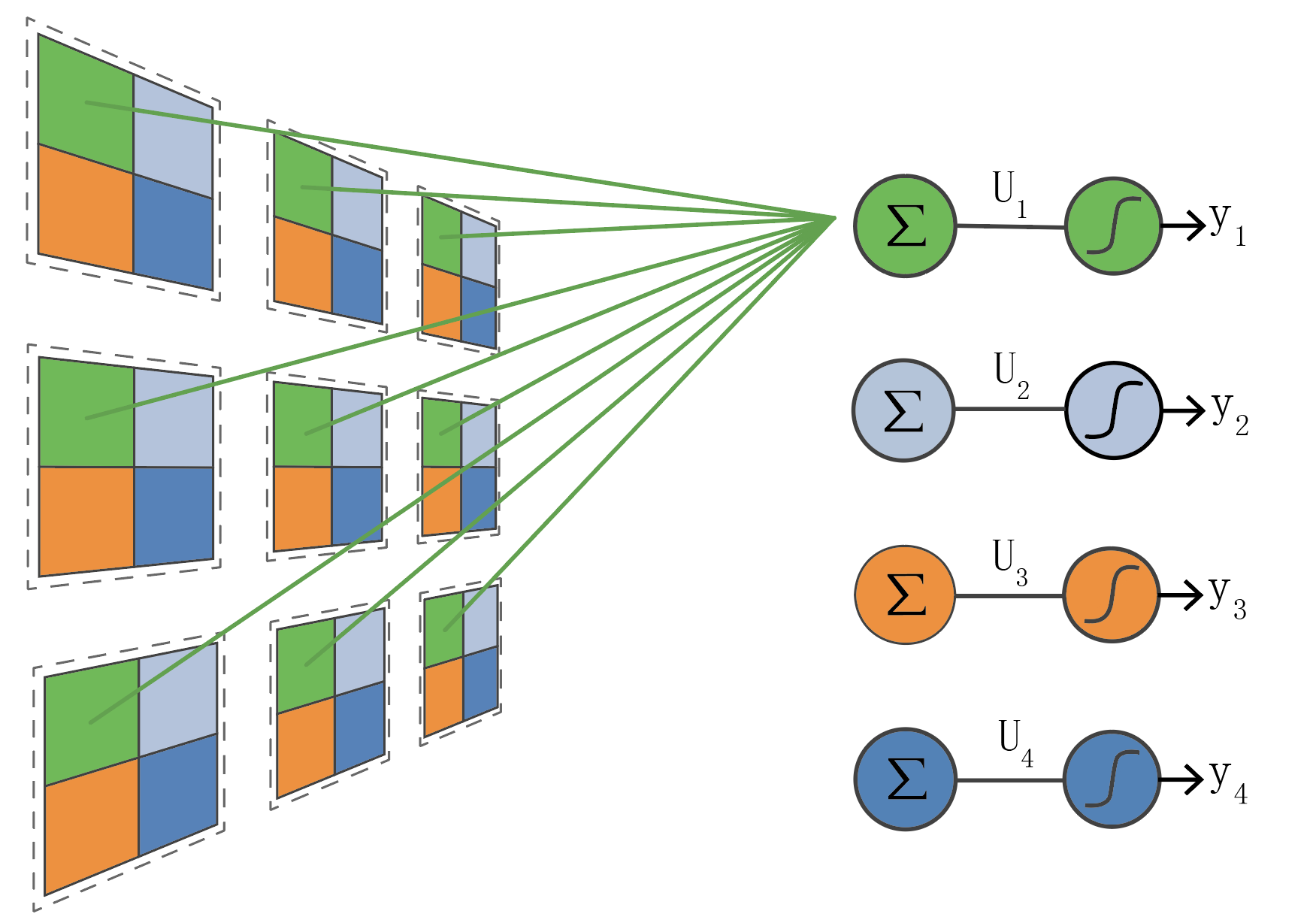}
		\caption{}
		\label{Fig6a}
	\end{subfigure}
	\hfill % 用于添加空间，使两个子图在水平方向上分开
	\begin{subfigure}{0.9\columnwidth}
		\includegraphics[width=\linewidth]{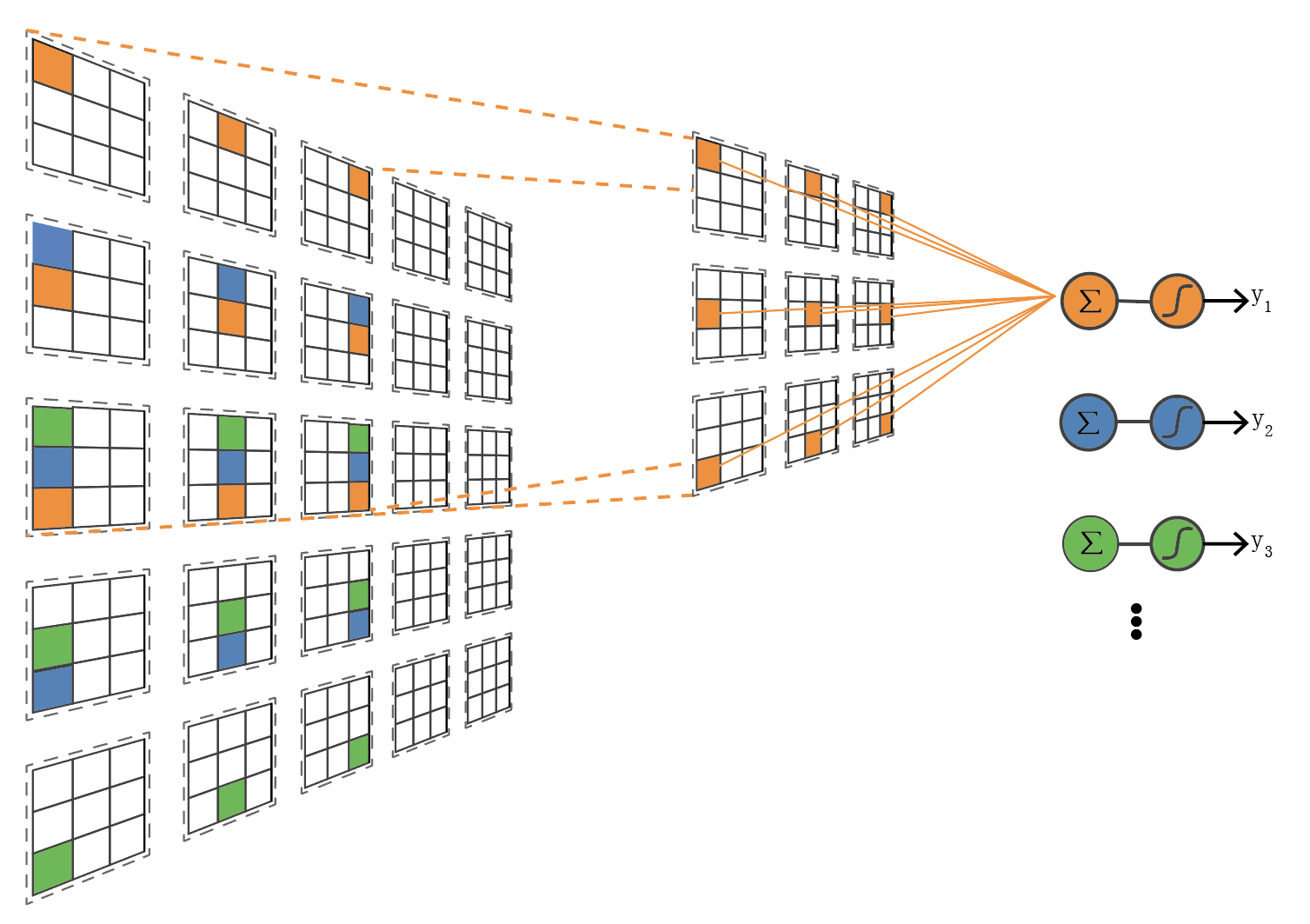} % 假设第二幅图的文件名为 Fig18b
		\caption{} % 第二幅图的标题
		\label{Fig6b}
	\end{subfigure}
	\caption{(a) Schematic of capacitive arrays for in-sense computation of fully connected layers. (b) Schematic of the convolutional windows with parallel computing.}
	\label{Fig6}
\end{figure*}

\begin{figure}[h]  % h 替换为 t 或者 ht 以获得更好的页面布局效果
	
	\centering
	\includegraphics[width=0.5\textwidth]{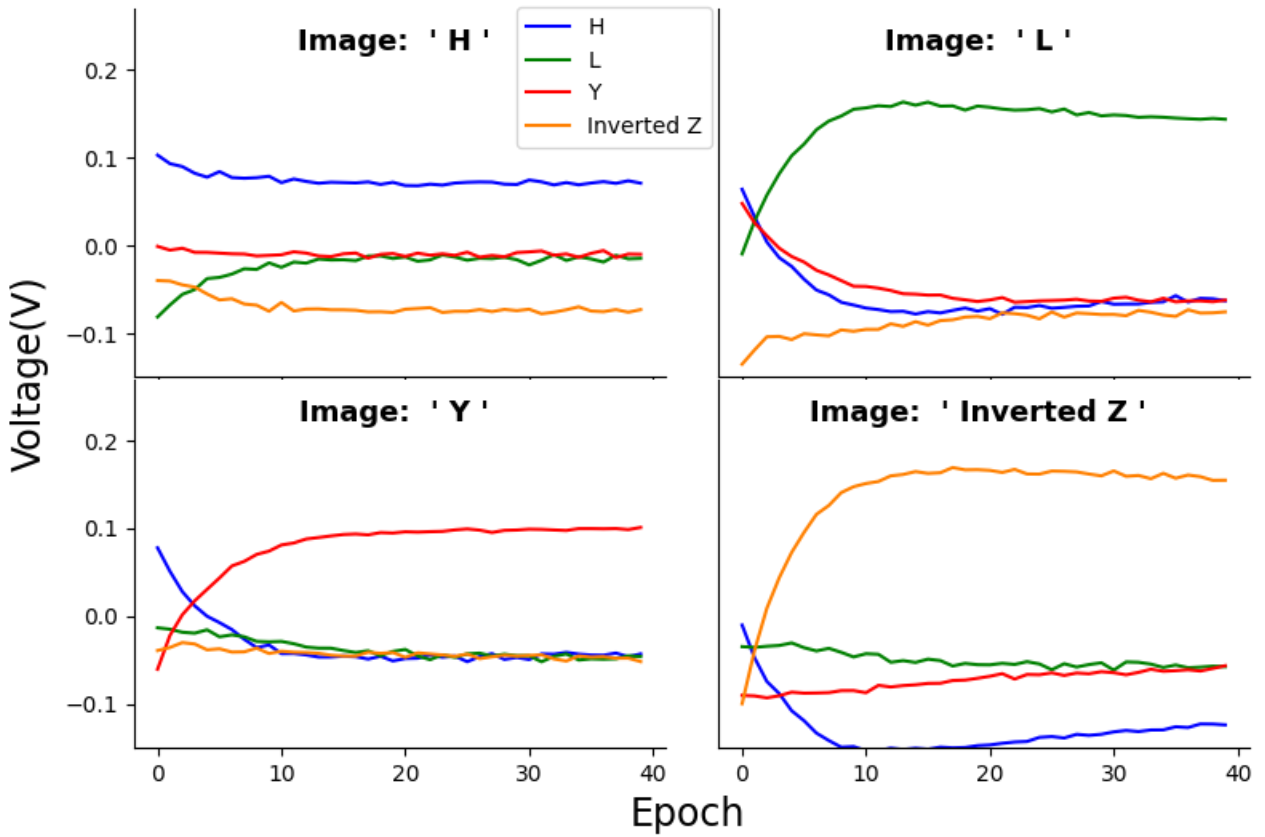}
	
	\caption{Measured average voltages for each epoch for a specific input letter, The curves with the highest voltage represent the recognition results.}
	\label{Fig7}

\end{figure}

 \subsection{FC NEURAL NETWORK OPERATIONS}
 \subsubsection{FC Classifier Operation}
Initially, a 3 × 3 capacitive sensor array is employed as a fully connected classifier to identify the letters H, L, Y, and inverted Z, as depicted in Fig.\ref{Fig3}(a). In ordere to explore the capability of the proposed circuits and algorithms, we set $C_0=72pf$ for sensing outside signal and use noise-free letters  with binary induced capacitance value as input, where the inside $C_{IH}=500pf$ and the outside $C_{IL}=16.77pf$. To augment the input, Gaussian noise with a root mean square value of 20$\%$ of the $C_{IH}$ or $C_{IL}$(totally labeled as $C_{Noise}$) is introduced. Consequently, the images are transformed into capacitive images with pixel values $ (C_{IH}+\delta C_{IH}) $ and $ (C_{IL}+\delta C_{IL}) $. Within a certain noise range, the fully connected classifier can categorize the images according to the labels shown in Fig.\ref{Fig5}.

In the sensor array shown in Fig.\ref{Fig6}(a), a capacitive pixel is composed of four subpixels formed by the circuits shown in Fig.\ref{Fig2}. Due to the proximity of the subpixels, it is assumed that $C_{I}$ sensed by each subpixel within a pixel is the same. The subpixels of a large capacitive pixel are designated as subpixels 1(green), 2(blue), 3(orange), and 4(deep blue). By connecting all subpixels of the same sequence from each large pixel, $U_1$, $U_2$, $U_3$, and $U_4$ can be output simultaneously. This parallel design can increase the computation speed of the circuit by four times.

\begin{figure}[t]  % h 替换为 t 或者 ht 以获得更好的页面布局效果
	\centering
	
	\begin{subfigure}[b]{0.23\textwidth}
		\centering
		\includegraphics[width=\textwidth]{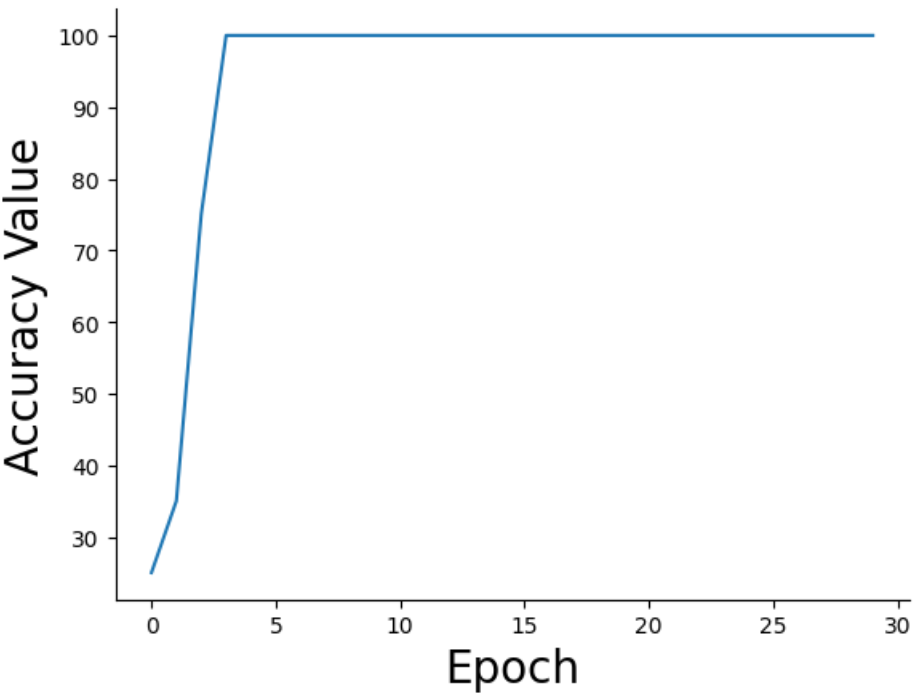} % 替换为您的图片文件名
		\caption{}
		\label{Fig8a}
	\end{subfigure}
	\hfill % 使用 hfill 而非 quad 以更好的分配空间
	\begin{subfigure}[b]{0.23\textwidth}
		\centering
		\includegraphics[width=\textwidth]{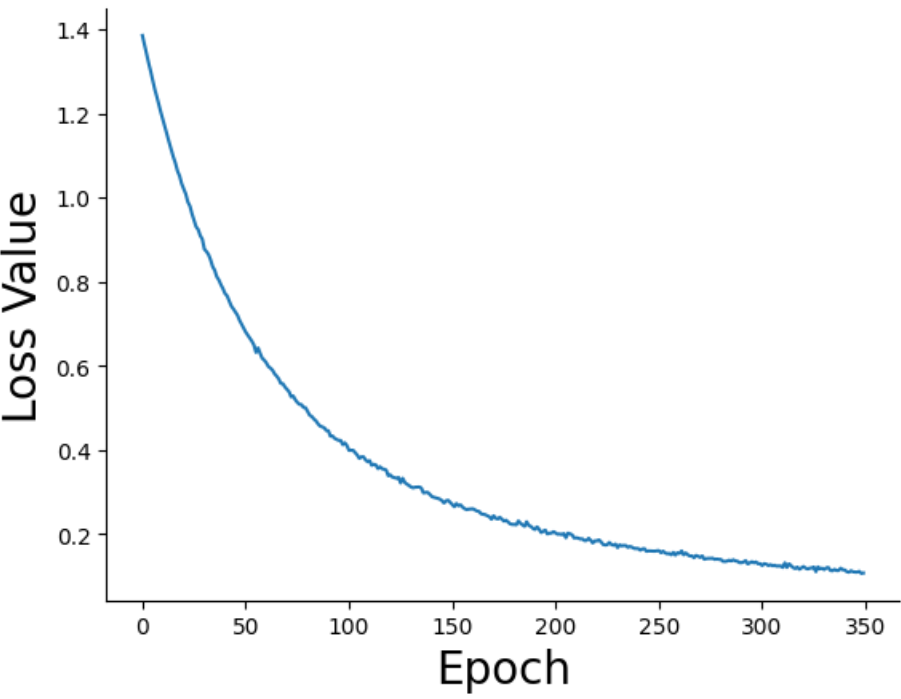} % 替换为您的第二个图片文件名
		\caption{}
		\label{Fig8b}
	\end{subfigure}
	\caption{(a) Accuracy and (b) loss of the FC classifier with noise $\sigma=C_{Noise}$.}
	\label{Fig8}
\end{figure}

Next, we introduce the training process of the fully connected classifier. We use HSPICE to construct a hardware circuit to replace the fully connected layer computation for training. Each epoch involves $S=20$ randomly generated capacitive letters. For each capacitive letter, we set up $M$ groups of $N$ voltages ($M=4$, $N=9$). The initial voltage weights are randomly distributed and normalized using the formula $V_{mn}^\prime=V_{mn}/\beta$ to ensure that the voltage weights are within the range of -1 to 1, where 
$\beta$ is the maximum absolute value of the voltage weights in that epoch. After biasing the capacitive sensor array with $V_{mn}$, four outputs $U_m$ can be obtained and activated in the digital domain using the softmax function. The loss is calculated using the cross-entropy function, and after each epoch, The weight $V_{mn}$ is updated to be $V_{mn}-(\alpha/S)\sum_{p}{\partial L/\partial V_{mn}}$, where  $L$  is the cross-entropy loss, $\alpha$ is the learning rate set to $\alpha=10$. After training for 350 epochs, the loss and accuracy curves of the fully connected classifier are shown in Fig.\ref{Fig8}(a) and (b). It can be observed that after 4 epochs of training, the accuracy quickly increases to 100$\%$, and the loss converges after 300 epochs. Additionally, as shown in Fig.\ref{Fig7}, the average output voltage corresponding to each letter is well separated after 4 epochs, with the maximum voltage corresponding to the input letter label.

   \subsubsection{FC Autoencoder Operation} 
\begin{figure*}[ht]  % 使用 ht 提供更灵活的放置
	\centering
	\begin{subfigure}[t]{0.23\textwidth}  % 增加宽度以更好地利用可用空间
		\centering
		\includegraphics[width=\textwidth]{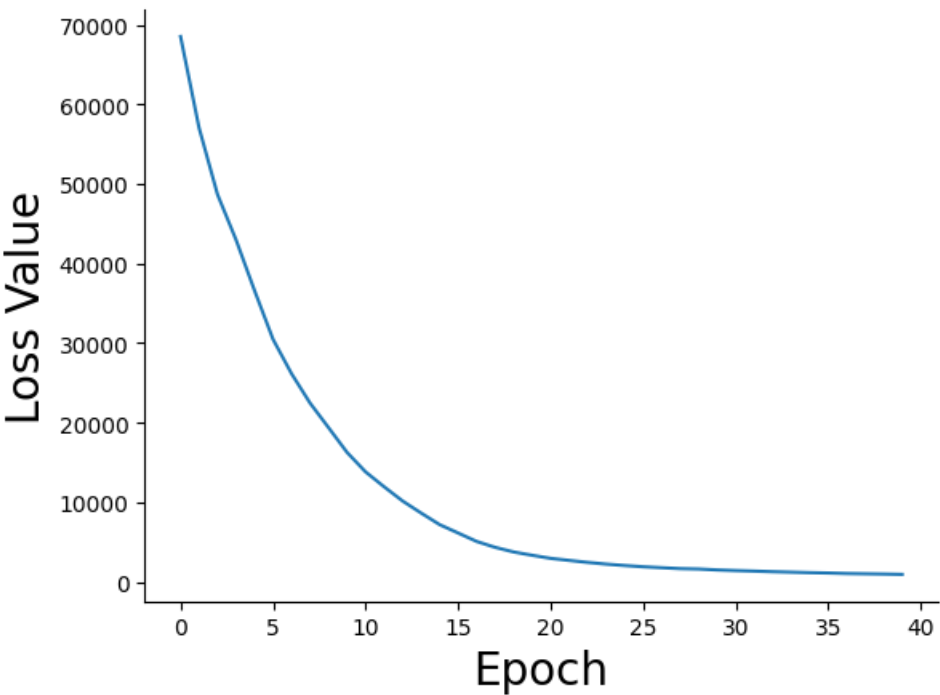} % 确保图像文件名正确
		\caption{}  % 添加具体的标题
		\label{Fig9a}
	\end{subfigure}
	%\hfill  % 确保使用 hfill 来均匀分配子图之间的空间
	\begin{subfigure}[t]{0.68\textwidth}  % 增加宽度以更好地利用可用空间
		\centering
		\includegraphics[width=\textwidth]{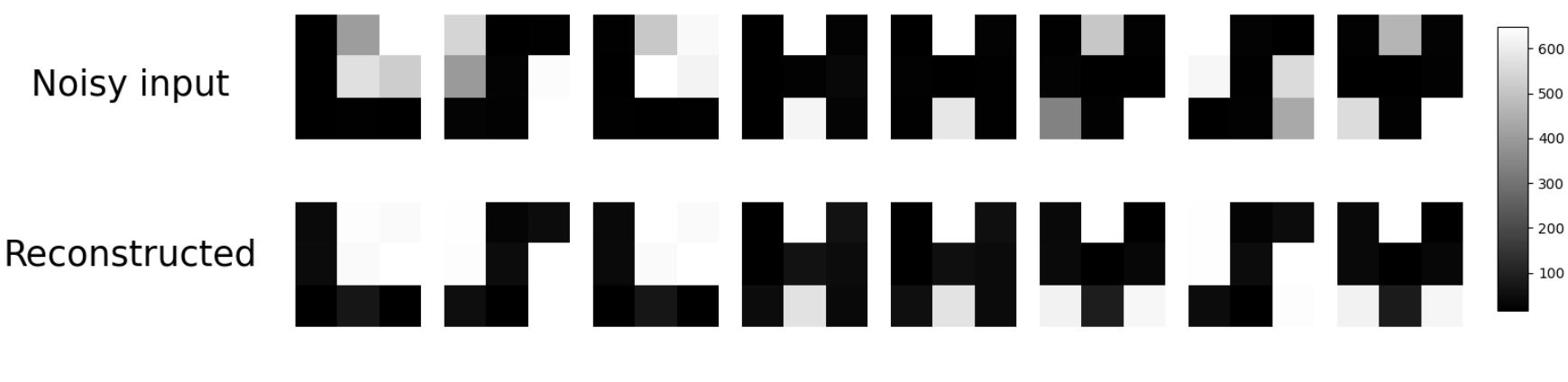} % 确保图像文件名正确
		\caption{}  % 添加具体的标题
		\label{Fig9b}
	\end{subfigure}
	
	\caption{(a) Convergence curves and (b) image reconstruction results for fully connected autoencoder with noise $\sigma=C_{Noise}$.}
	\label{Fig9}
\end{figure*}

\begin{figure}[t]  % h 替换为 t 或者 ht 以获得更好的页面布局效果
	
	\centering
	\includegraphics[width=0.5\textwidth]{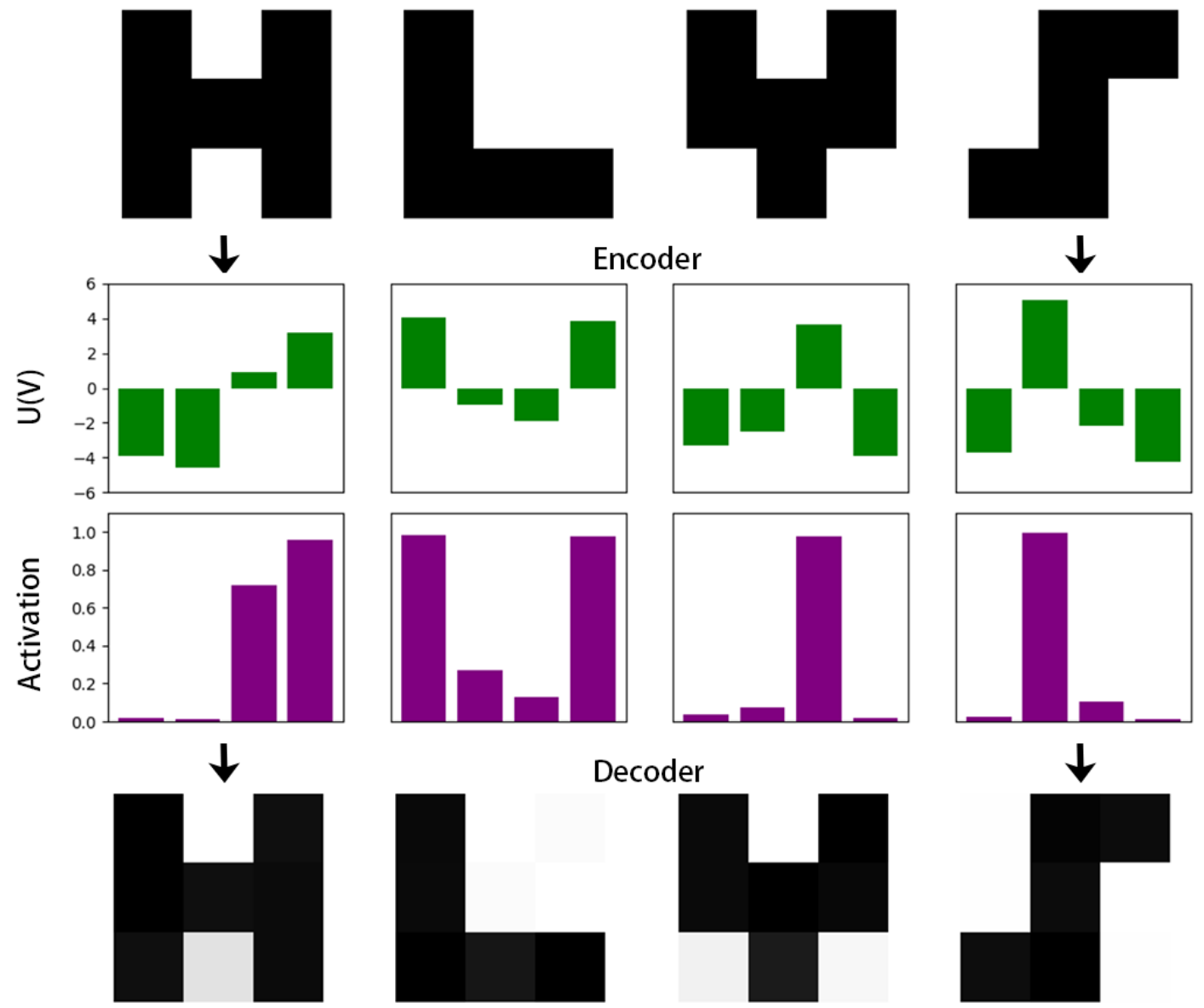}
	
	\caption{Processes of encoding and decoding for 4 noise-free letters. From top
		to bottom, the pictures correspond to the input of letters, encoding
		as voltage, activation values, and reconstructed letters at the
		bottom.}
	\label{Fig10}

\end{figure}

Now, we operate the capacitive sensor array as an FC autoencoder, as shown in the Fig.\ref{Fig3}(b). Similar to the FC classifier, we randomly input 20 capacitive letters with Gaussian noise $\sigma=C_{Noise}$ in each epoch, and the voltage weights are initialized by a PyTorch uniform distribution. To ensure that activation functions could reach the value for the decoder and accelerate the model's convergence, $C_{n}$ is normalized to be $C_{nl,n}={(C}_n-C_{L})/{(C}_{H}-C_{L})$. After normalization, the fully connected calculation output $U_m=\sum_{n=1}^{N}C_{nl,n}\cdot\ V_n=((A-B)/C)$, where $A=\sum_{n=1}^{N}{C_{n}\cdot\ V_n}$, $B=C_{L}\sum_{n=1}^{N}V_n$ and ${C=C}_{H}-C_{L}$. This process is similar to that of the FC classifier. $A$ is calculated in the capacitive sensor array, and the result is sent to the digital domain. The remaining processing is performed entirely in the digital domain. $U_m$ is the input to the sigmoid function ${\varphi_m(U}_m)=1/(1+e^{-U_m})$ for encoding. Subsequently, during the decoding process, the fully connected layer performs decoding to obtain: $Z_n=\sum_{m=1}^{M}W_{mn}\cdot\ {\varphi_m(U}_m)$.Through the sigmoid activation layer, we obtain: $C_{nl,n}^\prime=1/(1+e^{-Z_n})$, The denormalization is then performed to reduce the input capacitance vector to get $C_n^\prime=C_{nl,n}^\prime\times({(C}_{H}-C_{L})\ +C_{L})$, and the reconstructed induced capacitance value is  ${C}_{I,n}^\prime={(C}_{n}^\prime \times72)/(72-C_{n}^\prime)$, then use the mean square loss function $L=1/N\sum_{n=1}^{N}{{(C}_{I,n}^\prime-C_{I,n})}^2$ After each epoch, the encoder's voltages $V_{mn}$ and the decoder's weights $W_{nm}$ are updated via backpropagation with a learning rate of $\alpha=0.0004$. In Fig.\ref{Fig9}(a), the loss decreases sharply during the first 10 epochs and converges after approximately 15 epochs. During encoding, each noisy letter is sequentially encoded into voltage and activation values, as shown in Fig.\ref{Fig10}. For each letter, the activation values form a unique code of 4-digit values. Subsequently, the capacitive images are clearly reconstructed through the decoder. Finally, we applied 8 randomly selected letters with input noise of $\sigma=C_{Noise}$ to the capacitive array for reconstruction. The reconstruction results showed correct letters with less noise, as depicted in Fig.\ref{Fig9}(b).

 \subsection{CNN CLASSIFIER OPERATION}
\begin{figure}[t]  % h 替换为 t 或者 ht 以获得更好的页面布局效果
 	\centering
 	
 	\begin{subfigure}[b]{0.23\textwidth}
 		\centering
 		\includegraphics[width=\textwidth]{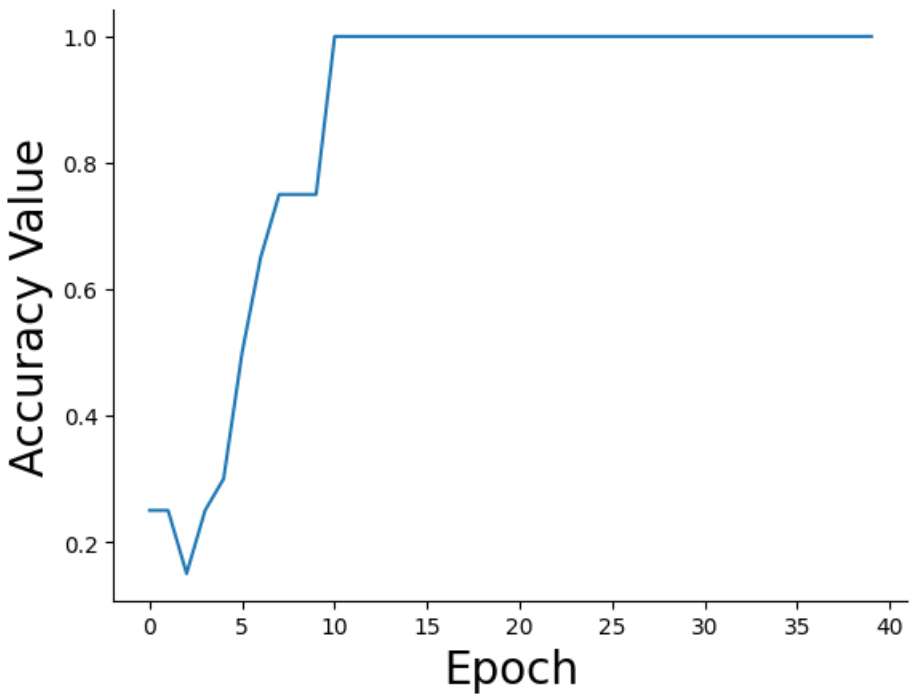} % 替换为您的图片文件名
 		\caption{}
 		\label{Fig11a}
 	\end{subfigure}
 	\hfill % 使用 hfill 而非 quad 以更好的分配空间
 	\begin{subfigure}[b]{0.23\textwidth}
 		\centering
 		\includegraphics[width=\textwidth]{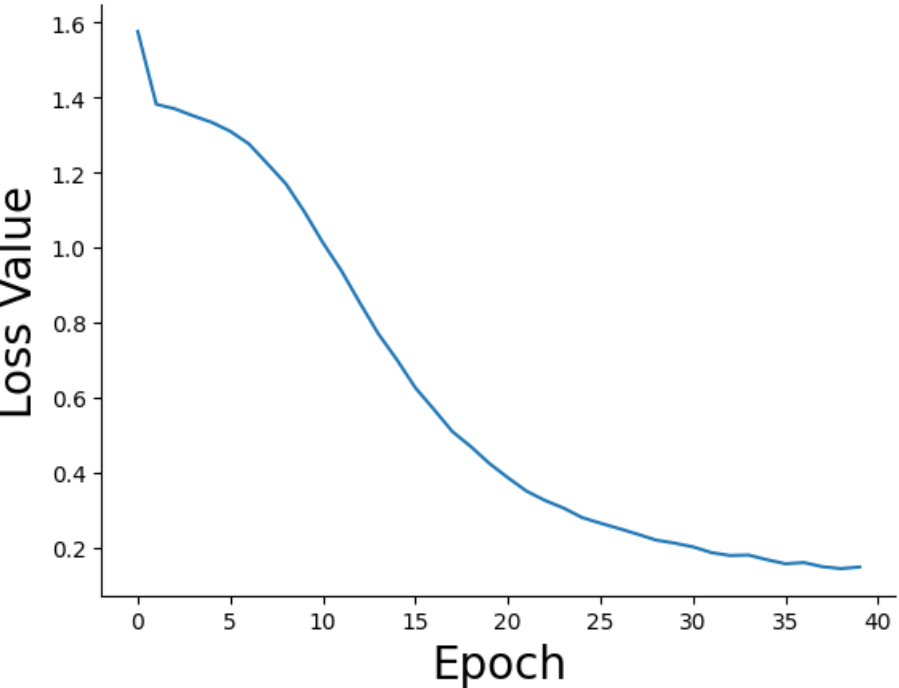} % 替换为您的第二个图片文件名
 		\caption{}
 		\label{Fig11b}
 	\end{subfigure}
 	\caption{(a) Accuracy and (b) loss of CNN classifier with noise $\sigma=C_{Noise}$.}
 	\label{Fig11}
 \end{figure}

Now, we use a 5×5 capacitive sensor array as a convolutional classifier to recognize the letters H, L, Y, and inverted Z, as shown in the Fig.\ref{Fig4}. First, we expand the 3×3 letter images shown in Figure 3 to 5×5. In each epoch, $S=20$ capacitive letter images with noise $\sigma=C_{Noise}$ are used as input. Unlike the fully connected classifier, the convolutional classifier computes the convolutional kernel in the sensor; 3×3 neighboring capacitive pixels form the convolutional kernel, and a total of 9 convolution operations are performed using the capacitive sensor array. The initial voltages/weights are randomly initialized. After in-sensor computing of the convolutional kernel, the output $U_m$ enters the digital domain and is activated by the Sigmoid function, then input to the fully connected layer. Finally, the output is fed into the softmax function for activation. Similar to the FC classifier, we use the cross-entropy loss function and backpropagation to update the voltage weights after each epoch. As shown in Fig.\ref{Fig11}(a) and (b), after 12 epochs, the accuracy reaches 100$\%$ and the loss reaches convergence after 30 epochs.

In the convolutional classifier, the capacitive sensor array we used is shown in Fig.\ref{Fig6}(b). Each capacitive pixel consists of nine sub-pixels. We achieved parallel computation of convolution windows through sub-pixel connections. In this 5×5 in-sensor computing array, three convolution windows in the vertical direction simultaneously compute and output voltages to the corresponding three ADCs. By horizontally moving the convolution windows twice, the computation results can be output completely. This approach can be extended to a $M \times N$ in-sensor computing array, where the convolution kernel results can be obtained in $(N-2)$ steps. In practical applications, such an array requires 9 DACs for providing weights and $M$ ADCs for outputting computation results. This design method is feasible for larger in-sensor computing arrays in practical applications\cite{ref7}.

 \begin{figure}[t!]  % h 替换为 t 或者 ht 以获得更好的页面布局效果
 	\centering
 	\begin{subfigure}[b]{0.23\textwidth}
 		\centering
 		\includegraphics[width=\textwidth]{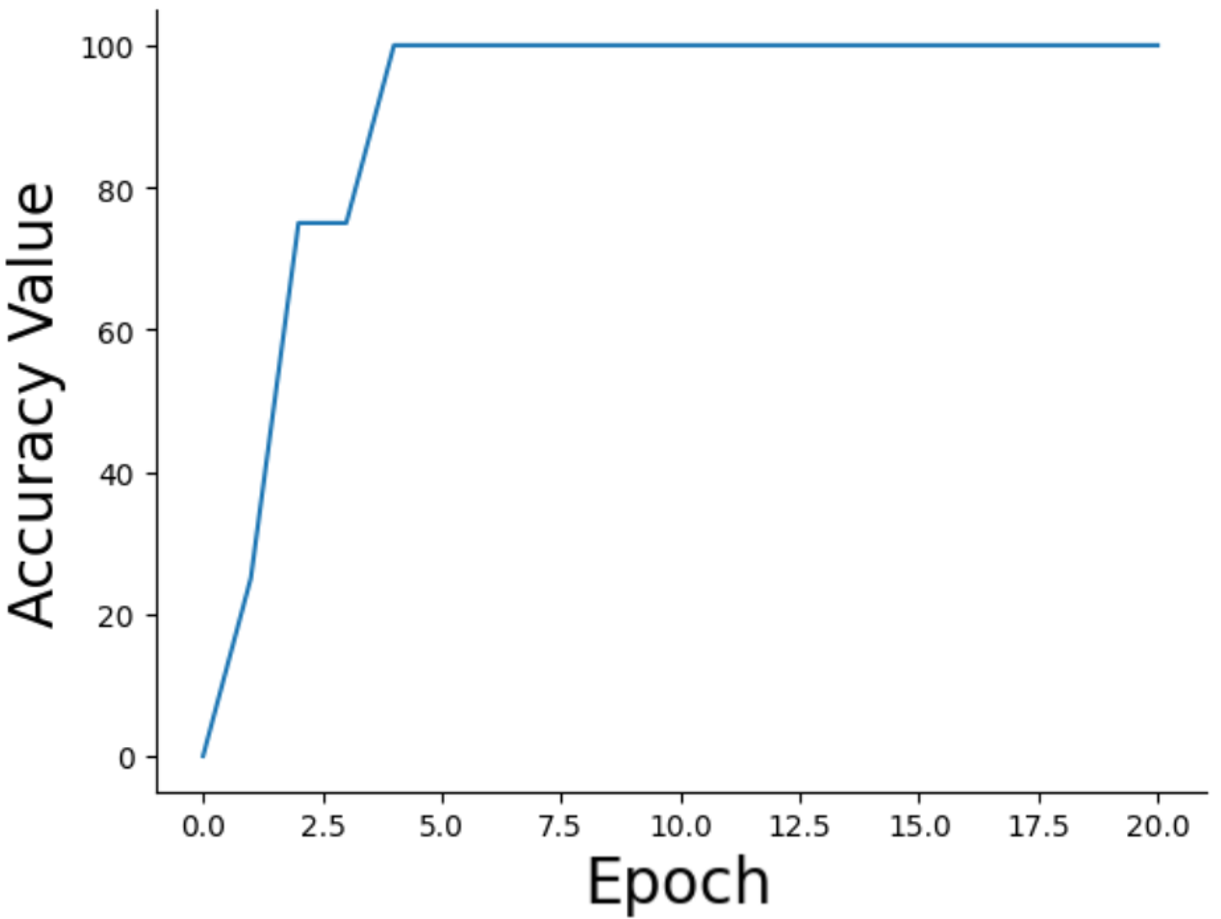} % 替换为您的图片文件名
 		\caption{}
 		\label{Fig12a}
 	\end{subfigure}
 	\hfill % 使用 hfill 而非 quad 以更好的分配空间
 	\begin{subfigure}[b]{0.23\textwidth}
 		\centering
 		\includegraphics[width=\textwidth]{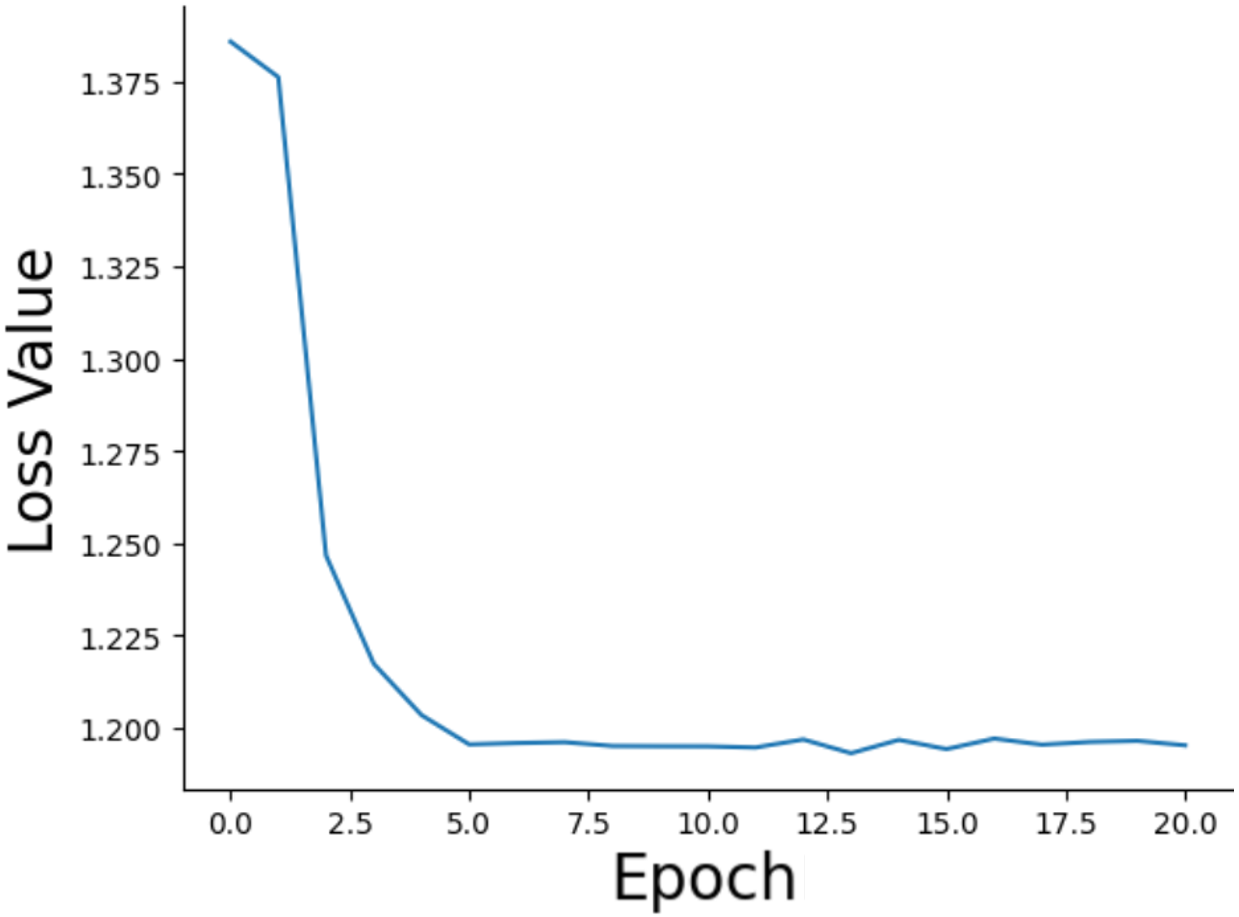} % 替换为您的第二个图片文件名
 		\caption{}
 		\label{Fig12b}
 	\end{subfigure}
 	\caption{(a) Accuracy and (b) loss convergence curves for the training of a fully connected classifier with voltage/weight binarization, when the noise is $\sigma=C_{Noise}$}
 	\label{Fig12}
 \end{figure}

\section{Evaluation and Discussion}

\begin{figure}[t]  % h 替换为 t 或者 ht 以获得更好的页面布局效果
	\centering
	\includegraphics[width=0.5\textwidth]{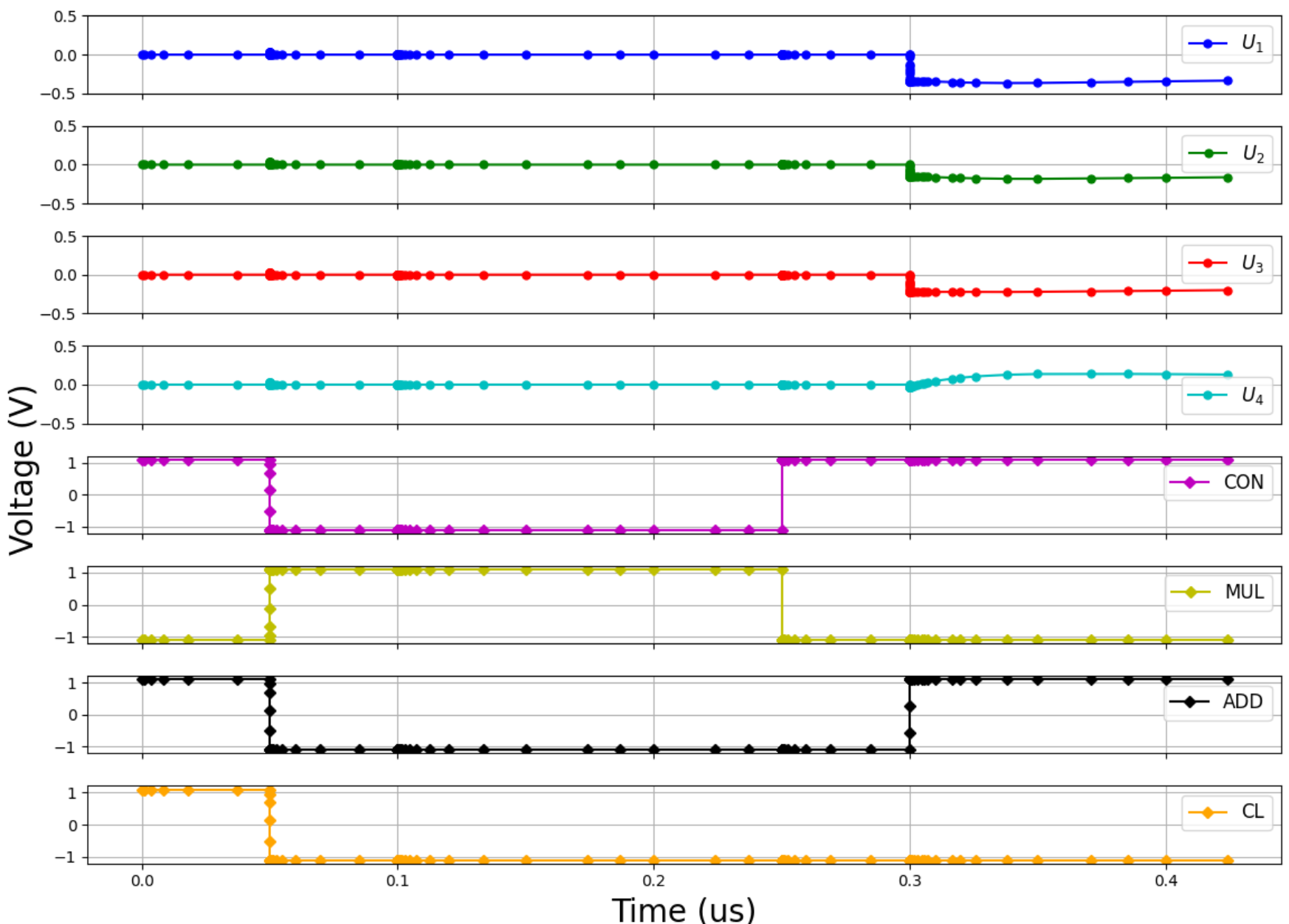}
	\caption{In-sensor computing capacitive array signal Waveform for "Inverted Z" input in a fully connected classifier.}
	\label{Fig13}
\end{figure}

In order to get the potential performance and power consumption of the proposed in-sensor computing circuit, we adopted a quantization method to binarize the voltage weights in the fully connected classifier to +1 and -1, using SRAM to construct the DAC module\cite{ref7}. 	The model converge rapidly, as shown in Fig.\ref{Fig12}(a) and (b). To evaluate performance, we measured and analyzed the delay of in-sensor computing using the aforementioned capacitive sensors. The computation time includes four operations performed by the circuit: charge clearing, capacitive charging, charge transfer, and charge summation. As shown in the Fig.\ref{Fig13}, the signal waveform diagram of a fully connected classifier with voltage/weight binarization is displayed when the input letter is "inverted Z". The output voltage of the fully connected layer stabilizes around $T\ \approx\ 0.35$ $\mu$s, where $U_4\ >\ 0$ and $U_1$ to $U_3\ <\ 0$, indicating a correct classification. Based on the HSPICE simulation results, the energy consumption of the FC layer computing is measured to be 0.9 $nJ$  for a correct classification.

\section{CONCLUSION}

In conclusion, we proposed In-sensor Computing ANN Capacitive Sensors, enabling rapid recognition and encoding of capacitive images. The effectiveness of these circuits and algorithms has been validated through HSPICE simulations. The proposed circuit demonstrates scalability and excellent energy efficiency. These advantages enable the widespread application of our proposed circuit and algorithms in the Internet of Things (IoT).

\vfill

\end{document}